\addtocontents{toc}{\protect} 
\documentclass[preprint]{elsarticle}
\usepackage[left=3cm,right=3cm,top=3cm,bottom=3.5cm]{geometry}
\usepackage{amsmath}
\usepackage[utf8]{inputenc}
\usepackage[english]{babel}

\usepackage{lineno,hyperref}
\modulolinenumbers[5]

\usepackage{graphicx}
\usepackage{subcaption}
\usepackage{float}
\usepackage{natbib}
\newcommand{\doi}[1]{\href{http://dx.doi.org/#1}{\texttt{doi:#1}}}

\usepackage{booktabs}

\makeatletter
\def\ps@pprintTitle{%
  \let\@oddhead\@empty
  \let\@evenhead\@empty
  \def\@oddfoot{\reset@font\hfil\thepage\hfil}
  \let\@evenfoot\@oddfoot
}
\makeatother

\newpage
\title{Forward propagation of a push through a row of people}

\author[1,2]{Sina Feldmann}
\ead{sina_feldmann@brown.edu}
\author[1]{Juliane Adrian\corref{cor1}} 
\ead{j.adrian@fz-juelich.de}

\cortext[cor1]{Corresponding author}
\address[1]{Institute for Advanced Simulation 7: Civil Safety Research, Forschungszentrum J\"ulich, J\"ulich, Germany}
\address[2]{Faculty of Architecture and Civil Engineering, University of Wuppertal, Wuppertal, Germany}

\begin{document}
\begin{frontmatter}

\begin{abstract}
Security plays a crucial role when it comes to planning large events such as concerts, sporting tournaments, pilgrims, or demonstrations.
Monitoring and controlling pedestrian dynamics can prevent dangerous situations from occurring. 
However, little is known about the specific factors that contribute to harmful situations. 
For example, the individual response of a person to external forces in dense crowds is not well studied. 
In order to address this gap in knowledge, we conducted a series of experiments to examine how a push propagates through a row of people and how it affects the participants. 
We recorded 2D head trajectories and 3D motion capturing data. 
To ensure that different trials can be compared to one another, we measured the force at the impact. 
We find that that the propagation distance as well as the propagation speed of the push are mainly functions of the strength of the push and in particular the latter depends on the initial arm posture of the pushed participants. 
Our results can contribute to a deeper understanding of the microscopic causes of macroscopic phenomena in large groups, and can be applied to inform models of pedestrian dynamics or validate them, ultimately improving crowd safety.
\end{abstract}

\end{frontmatter}

\section{Introduction}

Pedestrian dynamics play a crucial role in the safety of large events, such as concerts, sporting tournaments, and demonstrations. 
Crowded gatherings can easily lead to pushing and jostling, which can not only cause discomfort but also dangerous situations. 
A recent example of this is the Halloween celebrations in Seoul in 2022, in which over 150 people died and several dozen more were injured in a narrow, crowded street \cite{harrison_visual_2022}. 
To help prevent such tragedies, it is essential to better understand the dynamics of dense crowds and to identify potential measures to improve safety.
\\

When many people gather closely together, they interact and exchange forces, leading to various dynamics that influence the behaviour and movements of pedestrians as well as collective phenomena of the entire crowd. 
For example, turbulence can develop in dense crowds, causing pedestrian movements to become irregular and chaotic \cite{helbing_dynamics_2007, krausz_loveparade_2012}. 
This in turn can lead to a domino effect \cite{fruin_causes_1993, helbing_crowd_2012}, whereby a small disturbance starts a chain reaction spreading through the crowd.
Dense crowds can also exhibit 'density waves' where the density periodically increases and decreases \cite{bottinelli_can_2018}. 
In addition, collective movements where pedestrians move in a direction lateral to the desired direction have been studied in experiments \cite{garcimartin_redefining_2018, feliciani_systematic_2020}.
\\

One attempt to investigate these collective movements further is computer-based simulations based on pedestrian models.
Such models have the advantage of being cost-effective and allowing scenarios to be tested that cannot be studied in real experiments.
Modelling escape behaviour \cite{helbing_simulating_2000}, recreating a shock-wave through a crowd \cite{van_toll_extreme-density_2020} or incorporating crowding forces \cite{song_experiment_2019} are just a few examples.
However, these studies mainly manipulate model parameters to visualise specific situations in dense crowds without investigating real-life causes of such macroscopic phenomena.
\\

What can be macroscopically observed as transversal waves or large collective dynamics corresponds on a microscopic level to a transfer of momentum and forces acting between individuals.
Pushing is an example of a microscopic behaviour and can cause pedestrians to change their direction of motion or even fall.
In experiments, for example, pushing is used as a strategy to get to the target faster, which can result in a higher density \cite{adrian_crowds_2020}.
This often involves closing gaps, overtaking and pulling forward, which are assessed as pushing behaviour \cite{usten_pushing_2022}, \cite{alia_hybrid_2022}.
However, high densities in crowds can also be the cause for unintentional pushing because pedestrians touch each other due to the lack of space or moving in a confined space \cite{helbing_crowd_2012}.
\\

The way people react to an external force and how this force is propagated within a crowd have not yet been thoroughly investigated. 
Previous studies on this topic have mainly focused on pressure measurement at the wall and doorjamb during an evacuation of a crowd \cite{zuriguel_contact_2020}, pressure measurement between participants \cite{wang_study_2018, li_experimental_2020}, the development of a 'domino model' \cite{wang_modeling_2019}, or the definition of an individually perceived risk value \cite{wang_experimental_2020}. 
While these studies have provided valuable insights into contact forces and collision dynamics, they have limitations in terms of accounting for individual differences in pedestrian responses, and linking pressure data to motion data.
Small-scale experiments with one or two participants \cite{li_experimental_2021} analyse body postures and behaviours as a reaction to a force, but it is difficult to transfer these results to larger crowds.
\\

In our research, we want to contribute to a deeper knowledge of the processes of pushing by conducting a range of experiments.
The aim of the study was to determine the propagation of a push within a crowd.
For this purpose, head position, 3-dimensional motion data as well as pressure collected during small-scale experiments are analysed.
As a simplification, only one row of five people and one direction, i.e. forward movement are taken into account.
This allows us to break down reality to the smallest and simplest level and to investigate the effect of the push in an isolated way.
Thereby, the five people represent a small crowd occurring for example during waiting queues.
The overall objective of this study is to categorise pushes to compare the strength of the impact with the forward motion of the participants.
In the following, we will present our results in detail. 
We hope that our work can contribute to a deeper understanding of pushing in a crowd and provide useful information for the further development of pedestrian models and their application in practice.

\section{Methods}

The presented small-scale experiments were carried out at the research centre in Jülich in April 2021 as part of the EU-funded project CrowdDNA \cite{CrowdDNAProject2022}.

\subsection{Experimental setup}

\begin{figure}[htbp]
 \centering
\includegraphics[width=0.6\textwidth]{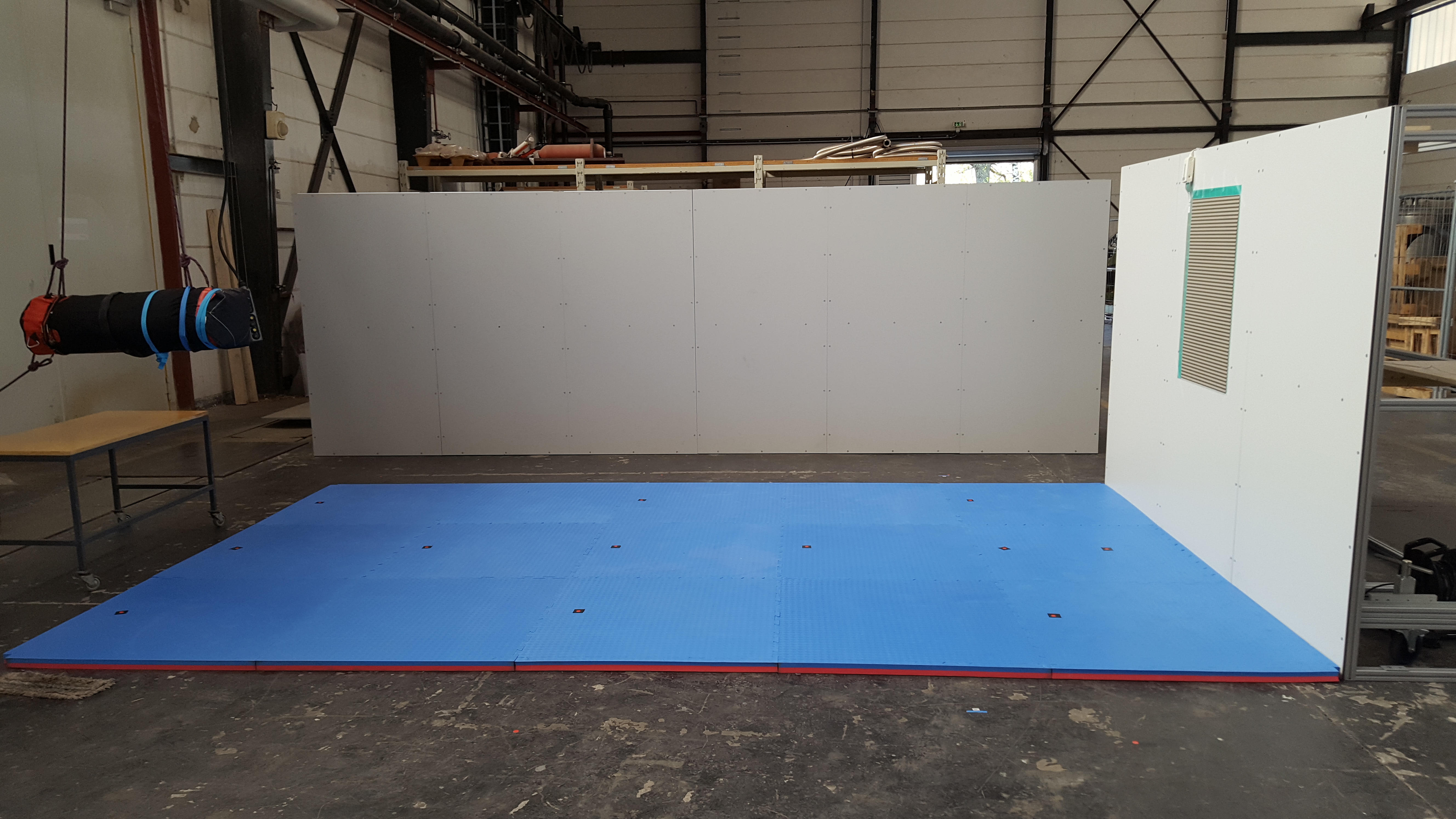}
\caption{The Experimental area covered with mats has a size of 3\,m $\times$ 5\,m. On the left side, a punching bag was suspended from the ceiling to push the participants forward. The right side of the area was limited by a solid wall.}
\label{fig:setup1}
\end{figure}

The experimental area of three meters by five meters (see Figure \ref{fig:setup1}) was covered with mats, that are also used in martial arts, to minimise injuries in case of falls.
One side of the area was limited by a solid wall, while on the other side  a punching bag, stabilised with a wooden plank, was suspended horizontally from the ceiling.

\subsection{Procedure and variations}

A total of 14 volunteers aged 19 - 55 years were recruited.
Prior to the experiments, each participant signed a written informed consent and assured that they felt physically fit enough to take part in the experiments.
The participants were informed beforehand about the procedure and possible risks. 
At any time, participants were able to stop the experiments or skip individual experimental trials without giving any reason or any disadvantage.
\\

In order to push the participants the punching bag was manually moved forward by the same person in all trials. 
Therefore, preliminary tests were carried out, in which pushes were classified into three categories: weak, medium and strong. 
To this end, four different participants were pushed individually determining the intensity of the pushes based on personal perception.
During this process, the person who moved the punching bag learned to push in the three defined categories. 
\\

Before each trial, five participants were lined up in front of the punching bag without touching it and facing the wall.
The conditions of the trial were read aloud and the distance between the participants were adjusted with the individual arm length. 
The person at the end of the line was pushed forward in a controlled manner. 

\begin{figure}[htbp]
 \centering
\includegraphics[width=0.9\textwidth]{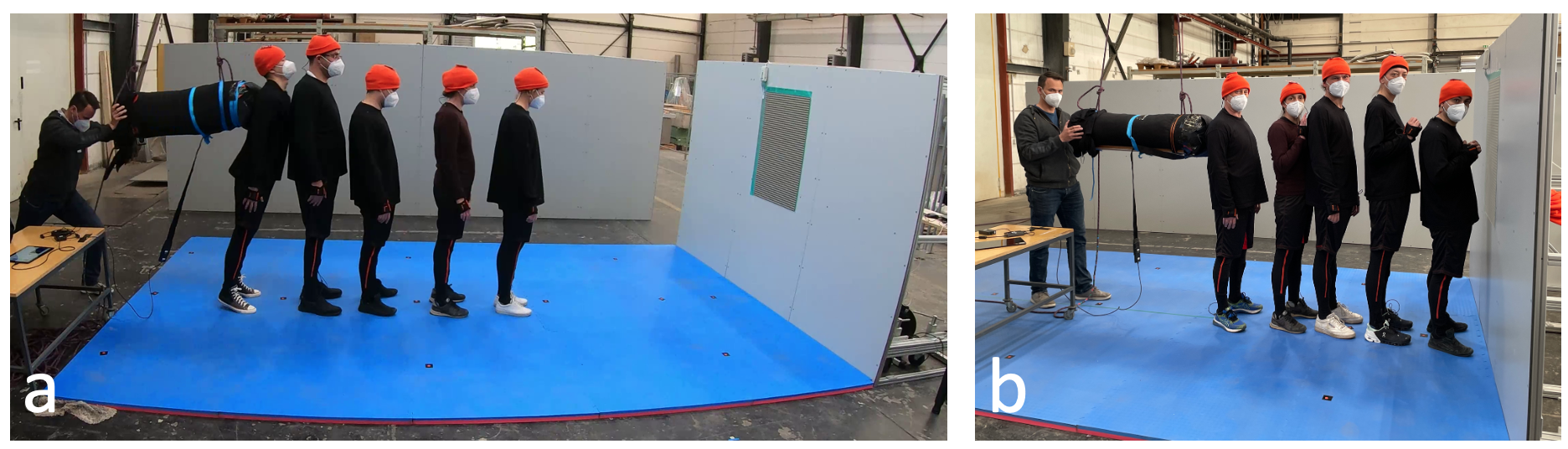}
\caption{Setup of the experiments (a) without wall and (b) with wall. Five people lined up in a queue and were pushed forward in a controlled manner.}
\label{fig:setup_participants1}
\end{figure}

The intensity of the push, the height of the push, the initial inter-person distance, the initial arm position and the body posture were varied (see Table \ref{tab:variations}).
In addition, the participants were positioned either directly in front of the wall with the first person keeping the same inter-person distance to the wall (i.e. with boundary), or far enough away from the wall to allow the first person to move freely to the front (i.e. without boundary), as shown in Figure \ref{fig:setup_participants1}.
Each variation was carried out firstly with a weak push and then the intensity of the push increased from medium to strong.
Upon request some runs were omitted.

\begin{table}[htbp]
\centering
    \caption{Overview of variations. Only variations in bold are analysed further.}
    \label{tab:variations}
    \begin{tabular}{l|lll}
    	\toprule
        \multicolumn{1}{c|}{Parameter} &  \multicolumn{3}{c}{Variations} \\
         \midrule
         intensity of push & \textbf{weak} & \textbf{medium} & \textbf{strong} \\
         initial inter-person distance  & \textbf{arm} & \textbf{elbow} & \textbf{none} \\
         initial arm posture & \textbf{free} & \textbf{down} & \textbf{up} \\
         body posture & \textbf{body tension} & relaxed \\
         height of push & \textbf{shoulder} & lower back \\
         boundary & \textbf{none} & \textbf{wall} \\
    \end{tabular}
\end{table}

For further analysis, only experiments in which participants stood with tension in the body, i.e. feet were placed hip-width apart, and were pushed at shoulder height are considered.
The series of experiments with and without boundary are considered separately.
Within a series, the order of the participants remained the same, but the intensity of the push, initial inter-person distance, and the initial arm posture were varied.
In total, 42 pushes without wall and 55 pushes with wall are analysed.

Trials in which persons stood relaxed could not be carried out in a standardised way, because individual interpretations of this condition were possible. 
The person at the end of the line moved quite differently when pushed at the lower back rather than the shoulder, which made it difficult to compare these trials with each other.
In order to simplify the experiments, not all variations, e.g. arm postures or inter-person distances, were included in the case of standing relaxed or being pushed at lower back.
Therefore, these trials are only used for a qualitative description.

\subsection{Data sets}
During the experiments, various data sets were recorded.

\subsubsection{Video recordings}

\begin{figure}[htbp]
 \centering
\includegraphics[width=0.7\textwidth]{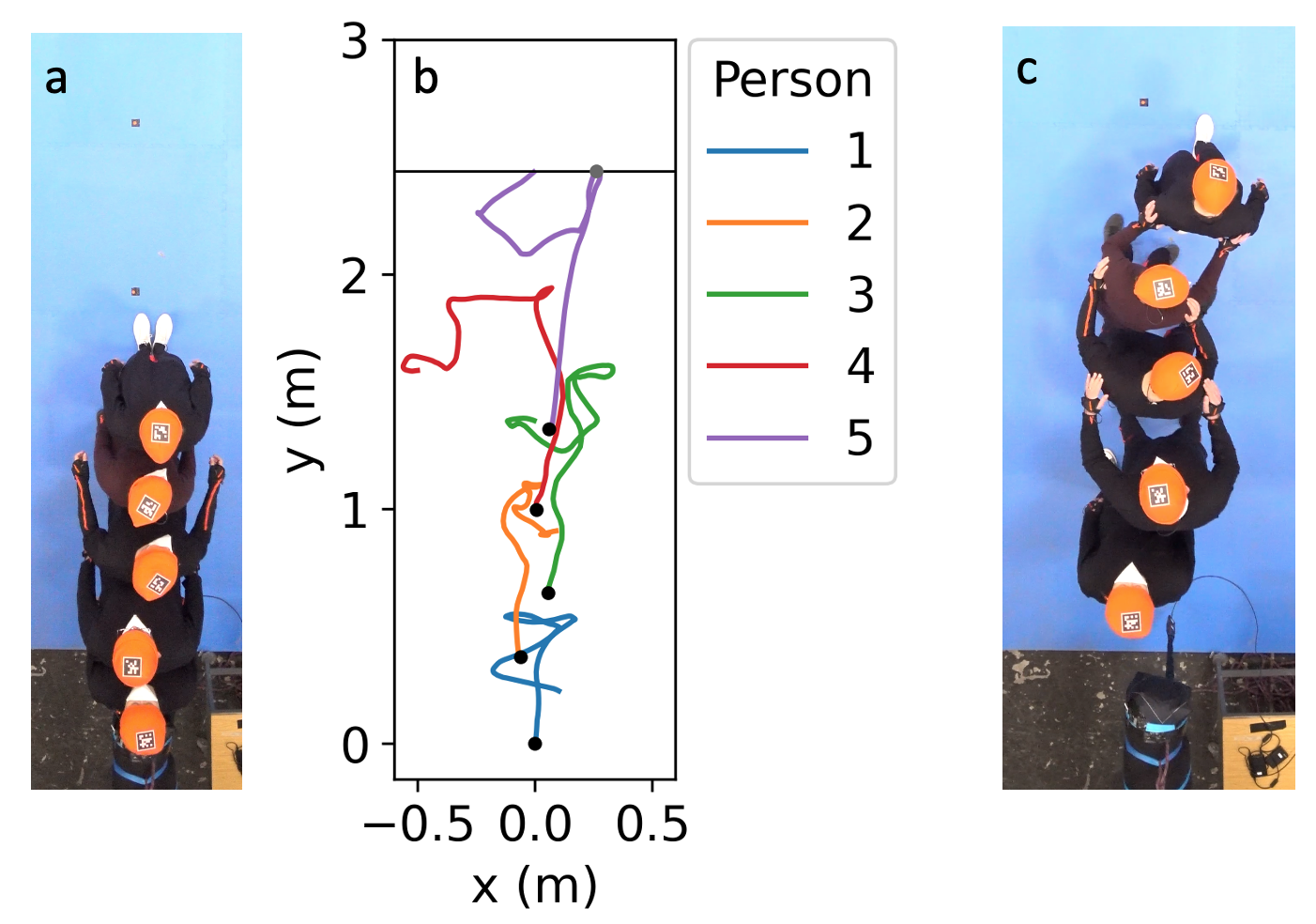}
\caption{(a) The snapshot from the overhead video recordings shows the participants at the beginning of the trial. (b) For all five participants, the start positions are represented as black dots and extracted head trajectories are plotted as coloured lines. It was pushed along the positive y-direction. (c) The snapshot from the overhead video recordings shows the position, when all participants have regained balance. This is set as end position for a push.}
\label{fig:trajectories1}
\end{figure}

The experiments were filmed with 25\,fps from a top-down as well as from a lateral perspective.
The side-view ensures a qualitative analysis of the trials.
During the experiments, each participant had to wear an orange hat with an attached Aruco-Code which is linked to individual characteristics such as body measurements and gender in an anonymised way.
These hats with codes are automatically detected and tracked using the video recordings from above in the Software PeTrack\cite{BOLTES2013127, boltes_maik_2021_5126562}.
The so derived individual head trajectories give information on the position within the experimental area (Figure \ref{fig:trajectories1}) and can be used to calculate other quantities, e.g. the velocity of the individual participants.

\subsubsection{3-dimensional inertial motion capturing}

All participants were dressed in a 3-dimensional motion capturing (MoCap) suit from Xsens \cite{schepers_xsens_2018}, which in turn was equipped with 17 inertial measurement units (IMU).
In order to record movements of each individual limb, the IMU sensors, measuring acceleration, the angular rate and the magnetic field strength, are placed on specific body segments, that can move separately from each other.
The advantage of this MoCap system, in contrast to other e.g. optical MoCap techniques, is that no full view of all body parts is necessary and therefore the Xsens suits can also be used in crowds.
\\

A calibration is required prior to the actual data recording.
For this purpose, detailed body measurements for all participants were taken.
By applying a biomechanical model in the MVN Analyze software, the orientation, position, velocity, acceleration, angular velocity and angular acceleration of every segment in addition to the joint angles and the position of the center of mass (CoM) are determined.
Each Xsens suit is connected wirelessly to the software MVN Analyze allowing a simultaneous start of the measurements.
All data is recorded with 240\,fps and stored on a local body pack in each suit.
To synchronise camera recordings with one another and with the motion data in time, a timecode generator \mbox{Tentacle Sync E} \cite{TentacleSyncGmbH} is used to impose time-codes on the measured data sets.
\\

\begin{figure}[htbp]
 \centering
\includegraphics[width=0.6\textwidth]{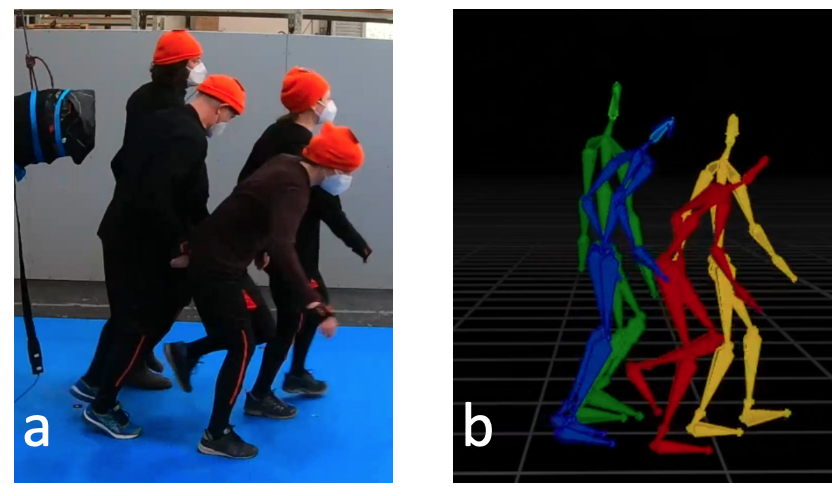}
\caption{(a) Participants wearing a 3D MoCap suit are pushed forward. (b) Combining 3D MoCap data with trajectories provides accurate positioning in the experimental area.}
\label{fig:MoCap1}
\end{figure}

However, the resulting 3D position data are considered for each individual separately and are not aligned to the global coordinate system of the experimental area, because the IMU sensors rely on relative measurements.
To ensure a correct spatial placement and orientation of all participants to each other, the 3D MoCap data are projected onto the camera trajectories (Figure \ref{fig:MoCap1}) by employing a hybrid tracking algorithm \cite{boltes_hybrid_2021}.
This allows the trajectories of the CoM to be used for further analysis of the motion initiated by a push, thereby neglecting large head movements.

\subsubsection{Pressure}

The pressure sensor Xsensor LX210:50.50.05 \cite{XsensorLX210:50.50.05} was attached to the punching bag and connected to the software Xsensor Pro V8.
Covering an area of 25.4\,cm\,$\times$\,25.4\,cm, 2500 measuring cells record pressure at a framerate of 34\,fps.
The pressure range is set to 0.14\,Ncm$^{-2}$ -- 10.3\,Ncm$^{-2}$, because the sensor has been pre-calibrated by the manufacturer.

In addition, the Pressure Mapping Sensor 5400N from TekScan \cite{TekscanPMS5400N} was mounted on the wall vertically and covered with Teflon foil.
This sensor measures the pressure with 1768 cells on an area of 57.8\,cm\,$\times$\,88.4\,cm at 90\,fps using the I-Scan software.
For a preliminary calibration, weights of 120\,kg and 185\,kg were placed on the horizontally lying sensor and mapped to pressure values with a sensitivity of S-30.
The sensitivity is a measure of how responsive the pressure sensor reacts.
We used a higher sensitivity than the default value to increase the resolution of the pressure measurements between the range of 0\,Ncm$^{-2}$ -- 8\,Ncm$^{-2}$.
\\

\begin{figure}[htbp]
 \centering
\includegraphics[width=0.8\textwidth]{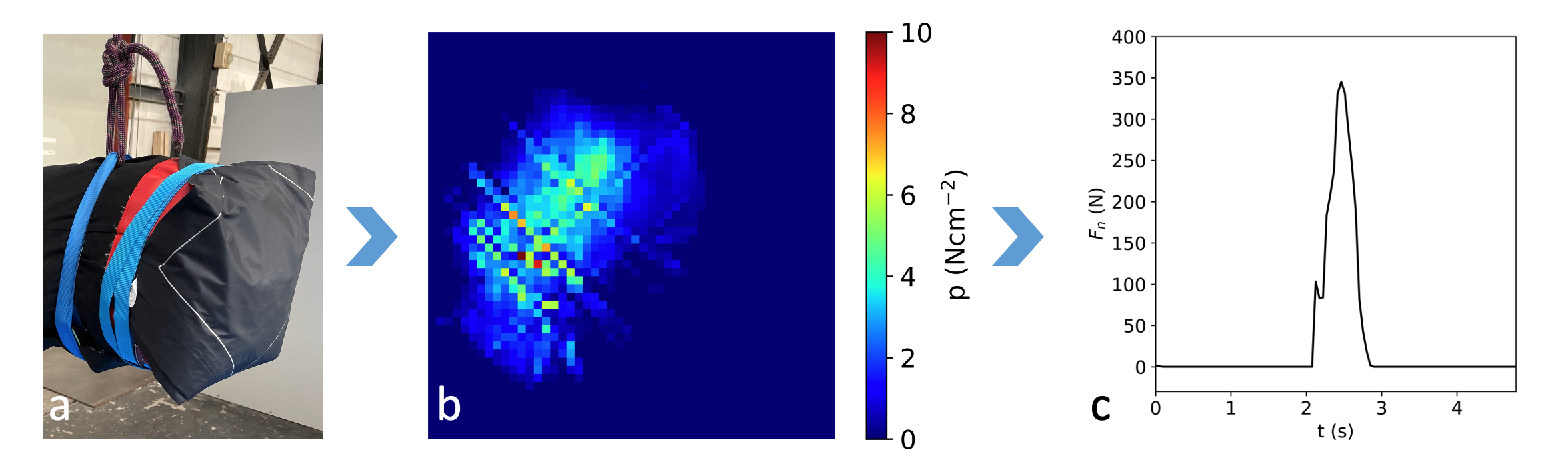}
\caption{(a) A pressure sensor measures the pushing intensities at the punching bag. (b) Spatial image of a single time frame recorded by the pressure sensor. (c) Time series of the normal force $F_n (t)$.}
\label{fig:pressure1}
\end{figure}

An instantaneous recording of the pressure sensor indicates the contact area of the punching bag with the first person pushed as well as the pressure on the sensor surface.
The pressure recordings $P$ are integrated over the area of the sensor $A$ for each time frame.
This results in a normal force $F_n $ acting on the back of the first person pushed.
The temporal development of this force provides information about other quantities such as the maximum force or the duration of a push (Figure \ref{fig:pressure1}).
The integral of $F_n $  over time determines an impulse of the push $J$.

$$ F_n (t) = \int_A P \, \text{d}A $$
$$ J = \int_t F_n (t) \, \text{d}t $$

\subsubsection{Accuracy}

In the coordinate system of the experimental area, the head trajectories have an error of 1.54\,cm and 1.20\,cm for the experiments with and without wall, respectively, as a result of the calibration in PeTrack and the consideration of the correct body heights for each participant.
In addition, there is an error of approximately 3\,cm because the correct position of the top of head, as also given in the 3D MoCap data, is important mainly for combining 3D MoCap data with trajectories.

The 3D MoCap data has a positional drift over time increasing the error of the positioning, which can be neglected because of the hybrid tracking algorithm.
Within one motion capturing suit, the accuracy of the individual limbs to each other have a dynamic error of 1\,$^\circ$ RMS.
The static accuracy of the trackers have an error for roll and pitch of 0.2\,$^\circ$ and for heading 0.5\,$^\circ$ \cite{Xsens2022}.
It should be noted that taking body measurements contribute an estimated error of 2\,cm to the biomechanical model.

The pressure sensor from Xsensor used at the punching bag has an accuracy of 5\,\% \cite{XsensorLX210:50.50.05} at full scale.
Based on the performed calibration, we assume that the accuracy of the pressor sensor from TekScan is 7\,\%.

\section{Results}
In the following, the trajectories of the CoM which result from the hybrid tracking algorithm are analysed because the head trajectories can show large movements.
Furthermore, we only consider movements in the pushing direction, i.e. in y-direction, because only a normal component can be measured at the pressure sensor.
This breaks down the experiment, which already represents a simplification of the reality e.g. a waiting queue, to the smallest and simplest level.
Thus, an isolated observation of how much of the push is really transmitted forward into the movement of participants without the influence of people standing around can be made.

\subsection{Pressure measurement}

One of the objectives of this study was to categorise the strength of the impact in order to achieve comparable pushes.
Considering that the strength of the impact is difficult to estimate for the manual pusher and the values cannot be read immediately at the pressure sensor, we aimed to obtain three different pushing intensities. 
According to the announcement, the punching bag was manually pushed forward in the categories: weak, medium and strong.
Absolut values for these three categories were not achieved and therefore the measured impulses are used for further analysis.
A detail comparison of the measured impulse of each push with the three categories  (weak, medium and strong) can be found in \ref{appendix:pressure}.
\\

Different impulses were measured for the three initial inter-person distances, which contradicts the expectation that the strength of a push only depends on the person who pushed.
This might be explained by the fact, that the pressure sensor responds to a resistance.
When the participants stand close to each other, the first person might already touch the second person while being pushed and thus increasing the resistance. 
It should be further noted, that the pusher could have unknowingly adjusted the pushes according to the resistance or experimental condition (with or without wall).
With the wall, the probability of injuries is more likely, so perhaps the pushing intensity was reduced for these trials.
Differences in the measured values for different initial inter-person distances as well as the conditions with and without wall can be seen in Figures \ref{fig:pressure_bag}\,b and \ref{fig:pressure_bag}\,a. 
\\

\begin{figure}[htbp]
 \centering
\includegraphics[width=0.6\textwidth]{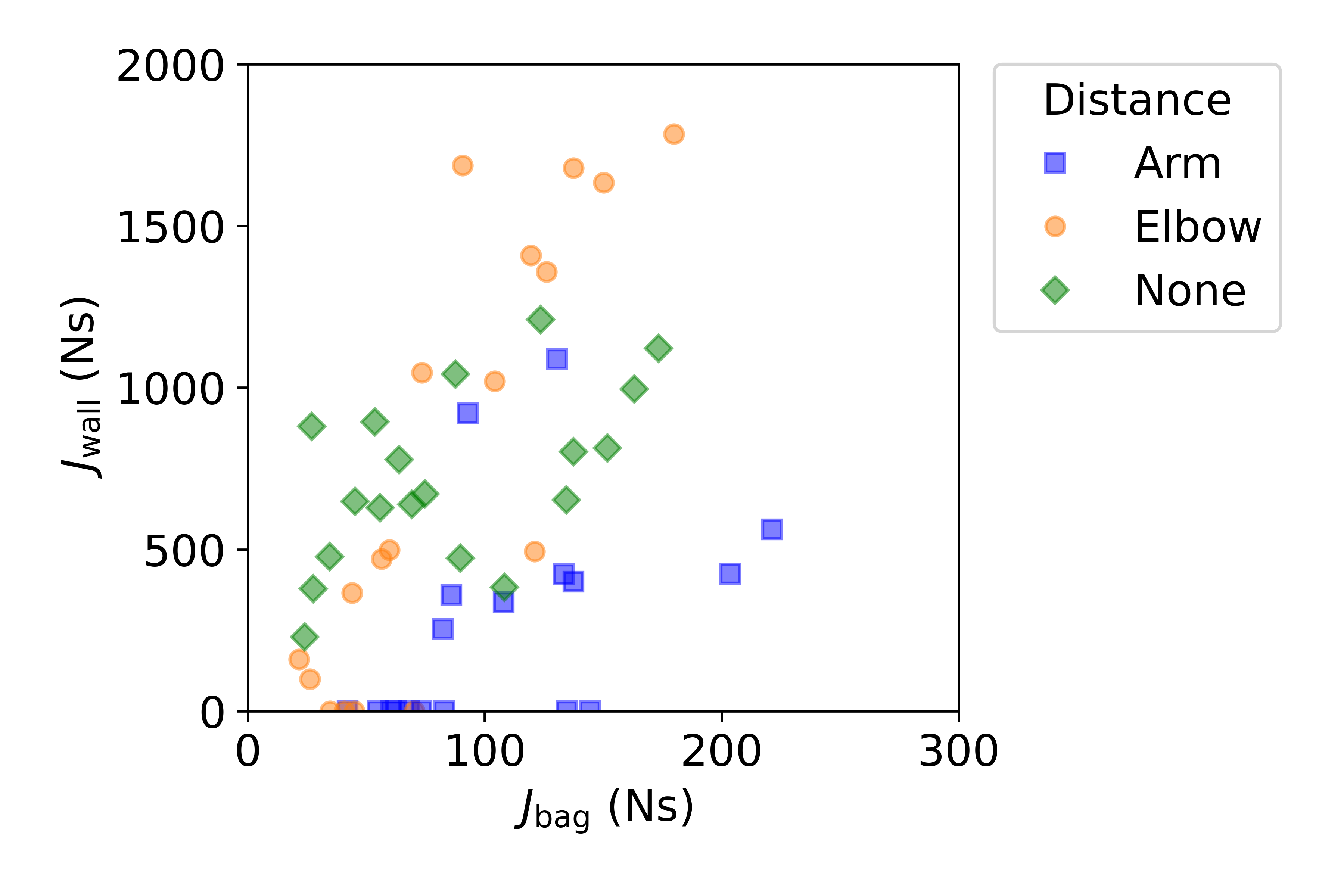}
\caption{Impulse measured at the wall compared to impulses measured at punching bag. Please note: Data recorded at the wall cannot be quantitatively compared to data recorded at the punching bag.}
\label{fig:pressure_wall}
\end{figure}

When the measured impulses on the wall are compared with the values measured on the punching bag, a certain correlation can be seen for different inter-person distances (Figure \ref{fig:pressure_wall}).
Here, the values on the wall for elbow distance are noticeable higher compared to no or arm distance.
This corresponds to the perception of the participants who rated the trials with elbow distance as most dangerous.

Furthermore, it has to be noted that the impulses measured at the wall are significantly higher than the impulses calculated at the punching bag.
The reason is that the two measurements were carried out using different sensors with different frame-rates and sensels sizes. 
The sensor on the punching bag has better surface resolution and a higher sensitivity while the sensor on the wall has a higher resolution in time.
Furthermore, the contact surfaces varied since it makes a difference if the back of a person is pushed with a deformable punching bag or bony elbows of a person touch a solid wall.
Most importantly, the contact time on the punching bag is limited whereas the contact time on the wall can exceed several seconds.
This makes the comparison of the data from the two pressure sensors difficult (Figure \ref{fig:pressure_wall}).
However, the data from the same sensor can be compared for different trials.

\subsection{Propagation distance of a push}

\begin{figure}[htbp]
  \centering
\includegraphics[width=\textwidth]{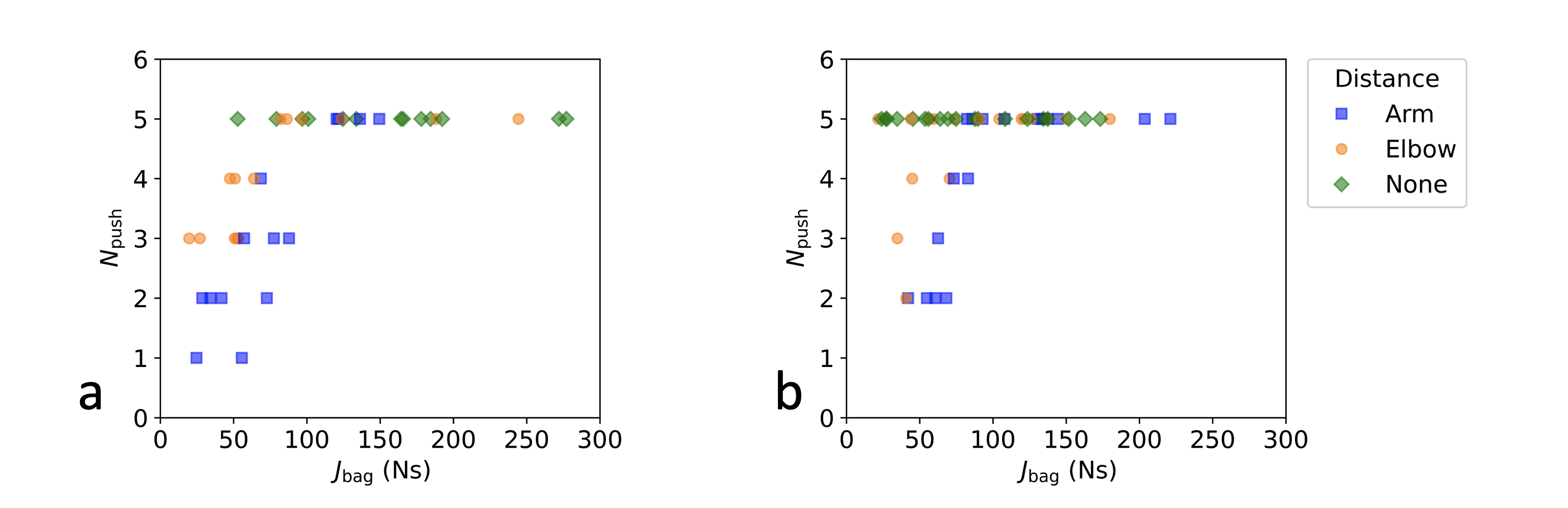}
\caption{Number of persons affected by the push in relation to the impulse at the punching bag for (a) experiments without wall and (b) experiments with wall. If the initial inter-person distance is small and the impulse is large, more people move forward due to the push.}
\label{fig:propagation_number}
\end{figure}

An analysis is made of how far a push propagates through a row of five people.
First, the number of participants who were affected by the push, i.e. who moved forward, is counted (Figure \ref{fig:propagation_number}).

The number of people moving forward due to the push depends on the initial distance of the participants to each other and on the strength of the push.
At no distance, all five persons are affected by the push regardless of the magnitude of the impulse.
At elbow distance, all participants move forward when the exerted impulse reaches over 75\,Ns. 
At arm distance, the push propagates through the entire row of people starting at a value of 90\,Ns.

To define a propagation distance of the push, the distance between the initial position of the person standing at the punching bag and the end position of the last person moving forward is calculated.
This takes two factors into account: The number of persons affected by the push (Figure \ref{fig:propagation_number}) as well as the distance the last person needed to regain balance.
In doing so, several steps could have been taken forward (Figure \ref{fig:trajectories1}\,c).
As an example, Figure \ref{fig:trajectories1}\,b indicates the distance of the push by a black horizontal line.

 \begin{figure}[htbp]
  \centering
\includegraphics[width=\textwidth]{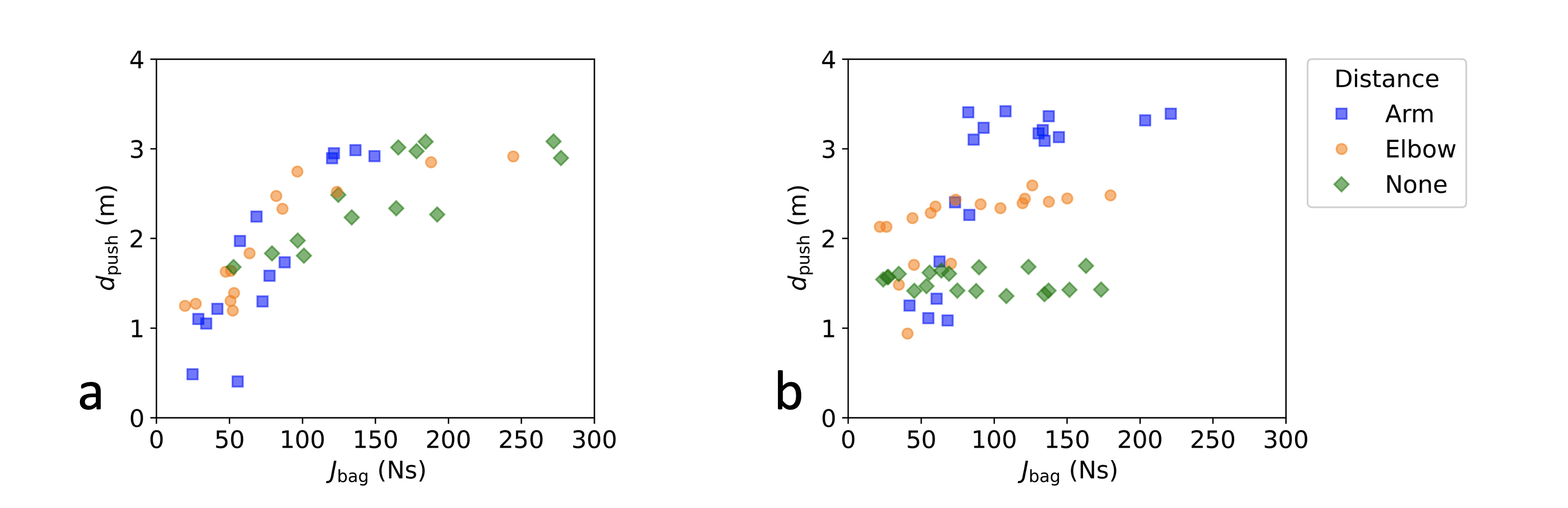}
\caption{Propagation distance of the push for (a) experiments without wall and (b) experiments with wall. (a) The propagation distance increases with stronger pushes before reaching a maximum. (b) The propagation distance is often the same as the distance to the wall.}
\label{fig:propagation_distance}
\end{figure}

For the experiments without wall, there is a linear relation between the propagation distance of the push and the impulse given into the system for impulses below 110\,Ns (Figure \ref{fig:propagation_distance}\,a).
Above 110\,Ns a maximum of 3\,m as the propagation distance is reached.
That could be related to the fact that the push passes through the entire row and the last person steps forward the same regardless of the intensity of the push.
Perhaps the wall at the end of the experimental area could also have an influence on this behaviour.
In the experiments with wall, the boundary is clearly visible (Figure \ref{fig:propagation_distance}\,b) and corresponds well to the findings in Figure \ref{fig:propagation_number}\,b.

\subsection{Propagation speed of a push}

\begin{figure}[htbp]
\centering
\includegraphics[width=\textwidth]{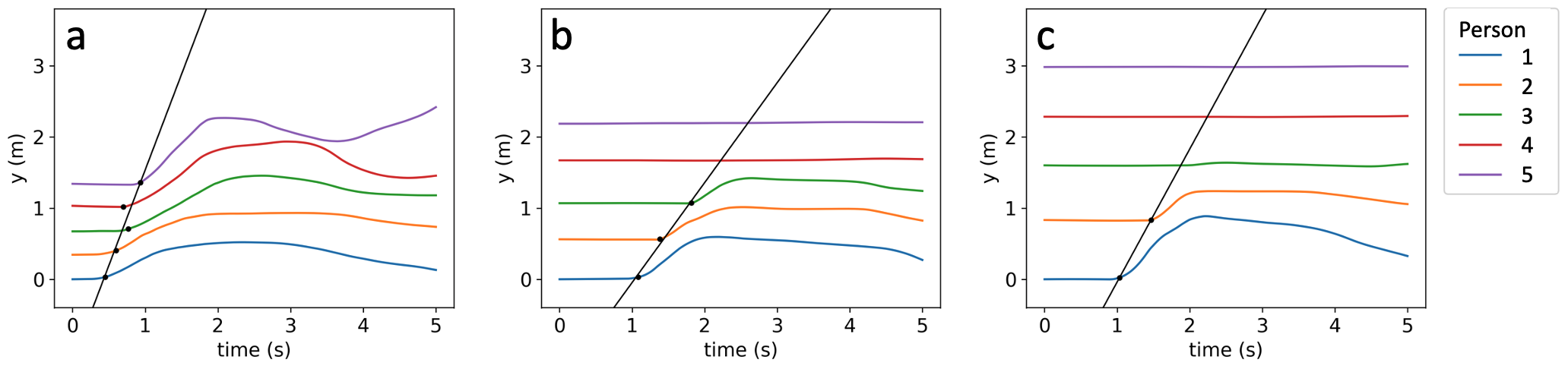}
\caption{Forward motion of CoM used to define a propagation speed of the push through the row of five persons for exemplary trials with different initial inter-person distances: (a) none, (b) elbow and (c) arm. The y-axis is aligned along the pushing direction. The regression lines (black) of the elbow points (black point) determine the propagation speed. When a person is not affected by the push, no elbow point is assigned.}
\label{fig:propagation_speed}
\end{figure}

For the definition of a speed at which the push propagates through the row, the forward motion of each participant is considered (Figure \ref{fig:propagation_speed}).
The elbow point of the y-t-curve (black dot) between a fixed start and end time of the push indicates the start time of the motion which corresponds to the time the push affects this person.
These points are aligned on a linear regression very well. 
Therefore, we define the propagation speed as the gradient of the regression line.

 \begin{figure}[htbp]
\centering
\includegraphics[width=\textwidth]{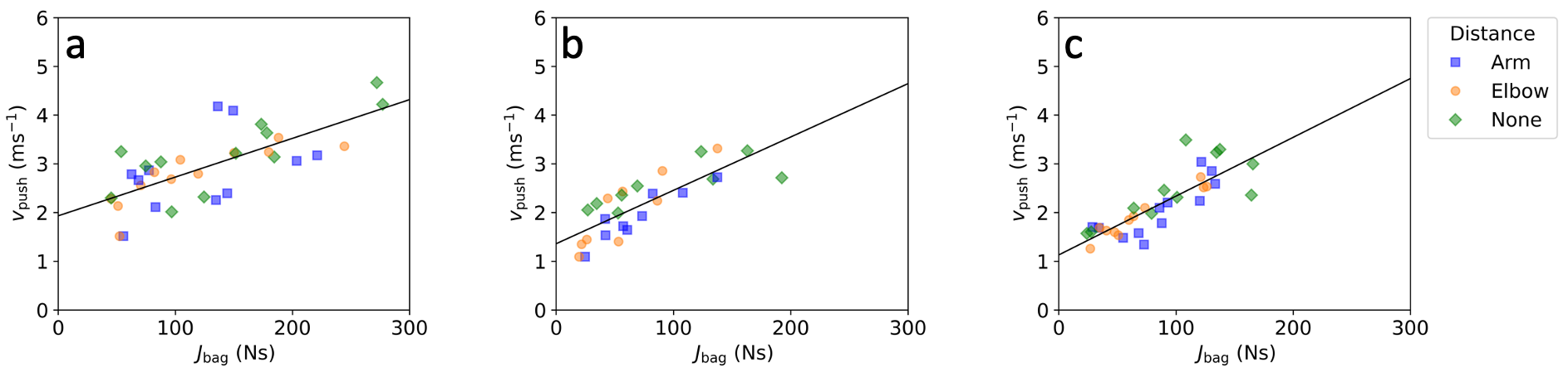}
\caption{Propagation speed of the push through a row of five persons for different initial arm positions: (a) up, (b) free, (c) down. In this case, trials with and without wall are considered together. A linear fit is indicated by a black line. The propagation speed increases with larger impulses.}
\label{fig:speed_impulse}
\end{figure}

Figure \ref{fig:speed_impulse} shows that the propagation speed depends on the intensity of the push. 
There is a linear relation between the propagation speed and the impulse of the push. 
However, the initial arm posture of the participants makes a difference, whereas inter-person distances have no effect on the propagation speed.
For the three initial arm postures (up, free, down), a multiple linear regression with interaction terms (moderation analysis) was performed in R.
The analysis did not show a significant difference between arms down and arms free.
Arms up has a significant effect on the propagation speed with  $\mathrm{p} < 0.001$ (between up and down) and $\mathrm{p} = 0.01$ (between up and free).
However, there is no significant interaction between arms up and the intensity of the push, as we found a p value of 0.055.
The fact that the result is just not significant could also be caused by the small sample size.
The complete output of the moderation analysis can be found in \ref{appendix:statistics1}.
We conclude that there are two equations for the propagation speed:

$$ v_{\text{push, up}} = 0.012 \, \frac{\mathrm{m}}{\mathrm{Ns}^2} \cdot J_{\mathrm{bag}} + 1.933 \, \frac{\mathrm{m}}{\mathrm{s}}$$
$$ v_{\text{push, down}} = 0.012 \, \frac{\mathrm{m}}{\mathrm{Ns}^2} \cdot J_{\mathrm{bag}} + 1.130 \, \frac{\mathrm{m}}{\mathrm{s}}$$

 \begin{figure}[htbp]
\centering
\includegraphics[width=0.6\textwidth]{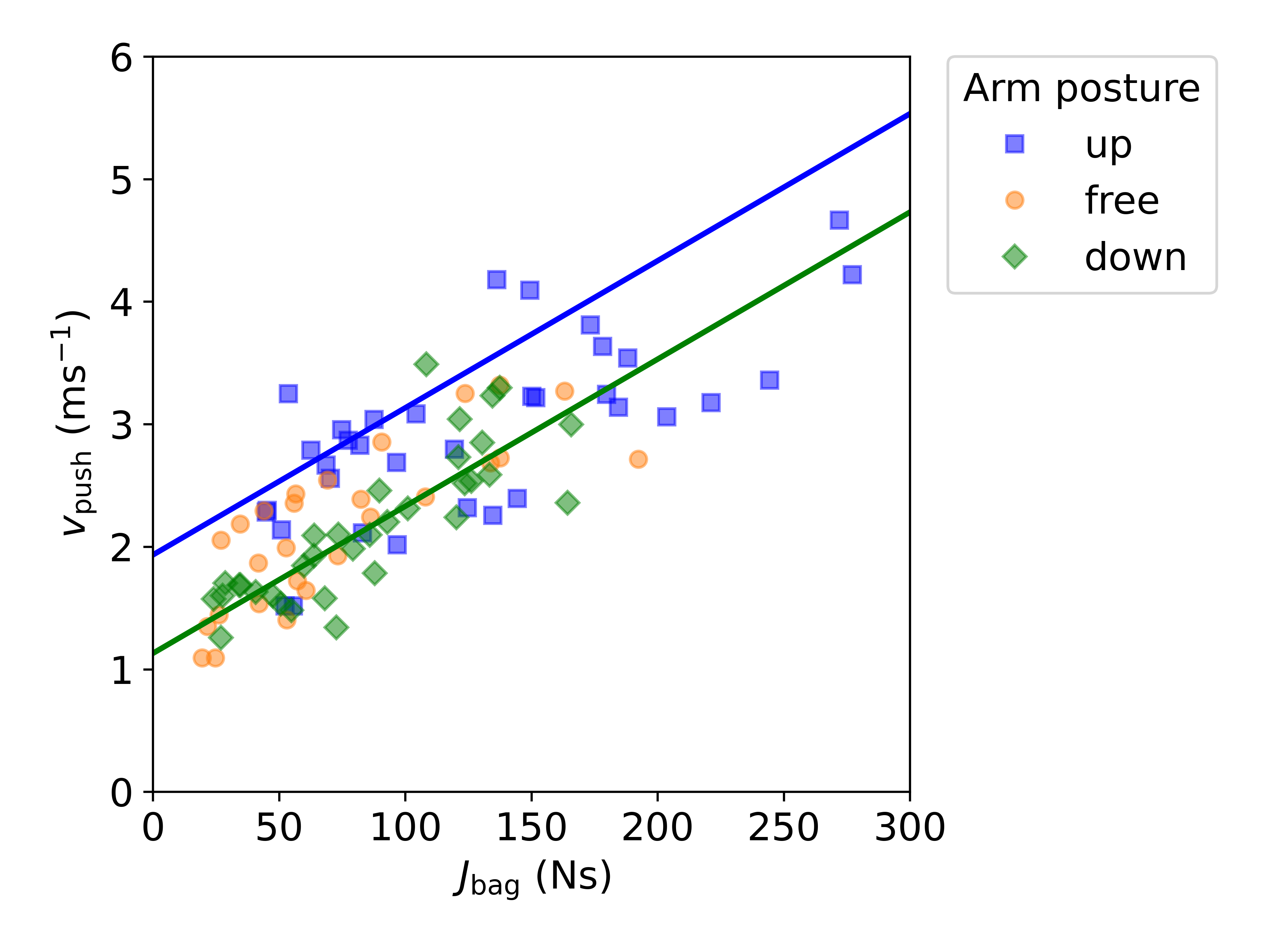}
\caption{Propagation speed of the push through a row of five persons for different initial arm postures (up, free, down). In this case, trials with and without wall are considered together. The propagation speed increases with larger impulses. For arms up or arms down/free, two linear fits can be obtained, shown as blue and green lines}
\label{fig:speed_impulse_comparison}
\end{figure}

A comparison of the two linear equations can be seen in Figure \ref{fig:speed_impulse_comparison}.
Through the effect of arms up there are different y-intercepts, but because we cannot assume an interaction the lines are parallel.
This means, a push propagates faster through the row if the participants keep their arms up from the start.
The results are consistent with the qualitative observation of the experiments, as people behaved very similarly with arms down and free. 
With arms up, the participants already touched each other from the start and were thus also able to pass on the push more quickly.
\\

\section{Discussion}

In this article, experiments were performed to investigate how pushes propagate through a crowd. 
For this purpose, a crowd was simplified as a row of five people, and only one pushing as well as one propagation direction were considered.
As soon as a person standing in a queue is pushed forward, the impact is passed on.
Thereby, it depends on the strength of the push, the initial inter-person distance and the initial arm posture how far or how fast the push propagates.
From a modelling perspective, this implies that the "dominoes" approach could be problematic, since in the naive concept of dominoes, the propagation of the push merely depends on the initial distance.
Therefore, it might be more appropriate to model pedestrians as inverted pendulums when they are exposed to an external force. 
\\

In our experiments we tried to classify the intensities of the pushes.
It is difficult to estimate forces in a crowd, because only the normal component of a force vector was measured with a pressure sensor and muscle forces of participants were neglected.
In addition, one person manually pushed the participants for all trials leading to a less reliable repeatability.
Based on different situations, the pushes can be adjusted.
For example a high resistance on the punching bag can intensify the measured impulses.
A wall at the end of the row, on the other hand, may reduce the pushing intensity, because the person manually pushing aims to avoid injuries at the wall.
\\

The trials with elbow distance were rated as being most dangerous by the participants.
These trials also often involved the highest impulses being measured at the wall, especially in the case of strong pushes.
The factors of space and time are probably important in explaining this.
It could be assumed that at elbow distance participants may have not enough space for their reaction (e.g. setting steps) and interact with one another very quickly, whereas at arm distance more space and time is available.
This can lead to more collision of the feet, which in turn increases the risk of tripping or even falling.
At no distance, the interaction between participants already starts before the push and they can be better aware of the available space for their own feet.
Another factor that may contribute to the increased danger at elbow distance is the speed with which interactions between participants occur. 
At elbow distance, each participant can accelerate more compared to no distance. 
But compared to arm distance, there might be hardly no time to slow down or restore their own balance before colliding with the person in front.
This would also result in more people reaching the wall at higher speed and thus increasing the measured impulses at the wall.
However, this is just an assumption and has to be further investigated for example by analysing the 3D MoCap data in more detail.
\\

The results presented in this article provide valuable information on propagation distance and speed of a push taking into account different body positions as well as pushing intensities.
This in turn can give insight into how people move one another, intentionally or unintentionally, and thereby help to better understand dynamics even in larger crowds.
These information can be used to validate and improve pedestrian models as well as to help assess risk levels and prevent potentially dangerous situations.
In our analysis, it is especially interesting to observe, that the initial inter-person distance has no effect on the propagation speed.
This means that the propagation speed is not dependent on the quasi 1D density.
However, just applying this concept to a larger crowd with 2D density seems a bit  tricky.
This analysis should be investigated on a larger sample size to draw a statistically significant conclusion.
Besides, the propagation speed needs to be analysed for 2D densities, for example in experiments where several rows are standing next to each other. 
\\

Furthermore, other factors such as individual characteristics, body tension, and also preparedness can play an important role in how a push propagates.
People can respond in individual ways by choosing different strategies (ankle, hip, steps \cite{winter_human_1995}, \cite{maki_control_2006}, \cite{tokur_review_2020}) to regain balance.
This could result in either absorbing or intensifying the impact.
Individual motion strategies and associated factors that influence the transmission will be investigated in more detail in the future using the 3D motion data.
We will focus on the evaluation of the steps (step length, step width and number of steps), movements of the hips as well as pendulum movements of the upper body.
\\

The small-scale experiments presented here examined a total of 97 pushes on eight different participants.
This is a small sample size that may not allow for a reliable statistical analysis.
Therefore, the proposed analysis will be applied to larger experiments including up to thirty participants.
So far, only a very simplified representation of the crowd in form of a row of people have been considered, which can limit the generalisability of the findings.
Our results can mainly be applied to real-world situations where people are standing along a line, e.g. a waiting queue as observed at the entrance of a concert.
\\

But in real-world situations, people are often distributed in a random and irregular order.
Especially in larger crowds, it is important to consider the arrangement of people in order to investigate the propagation of a push more accurately.
In upscaled experiments, positioning of the participants will be varied in order to obtain a more realistic representation of crowds.
The persons will stand either in a long row, in several rows next to each other or staggered as a group.
Thus, not only the propagation of the push to the front will be investigated, but also how the push propagates to the side and therefore including density into the analysis.
Furthermore, the analysis will focus on the extent to which the transmission of an impulse differs from how the participants were prepared.
\\
\\

\noindent \textbf{Authors contribution}\\
Sina Feldmann: Writing \textendash{} original draft, Visualization, Formal analysis, Data curation, Conceptualization. 
Juliane Adrian: Writing \textendash{} review \& editing, Supervision.
All authors agree to the publication of the manuscript. 
\\

\noindent \textbf{Declarations of interest} \\
The authors declare that they have no known competing financial
interests or personal relationships that could have appeared to
influence the work reported in this paper.
\\

\noindent \textbf{Data Availability}\\
The raw data from the experiments are freely accessible at the Pedestrian Dynamics Data Archive of the Research Centre J\"ulich,
\mbox{\doi{10.34735/ped.2022.2}}.
\\

\noindent \textbf{Acknowledgements}\\
We would like to thank Jernej {\v C}amernik, Marc Ernst, Helena Lügering, Armin Seyfried and Anna Sieben who supported us in the conception, planning and realisation of the experiments.
Furthermore, we are grateful to Alica Kandler, Maik Boltes, Ann Katrin Boomers and Tobias Schrödter for their help in setting up the experiments and curating the data.
\\

\noindent \textbf{Funding} \\
This research was supported by the European Unions Horizon 2020 research and innovation program within the project CrowdDNA [grant number 899739].
\\

\noindent \textbf{Ethical Review} \\
The experiments were approved by the ethics board of the University of Wuppertal in April 2021 (Reference: \mbox{MS/BBL 210409 Seyfried}).
\\

\bibliographystyle{cdbibstyle} 
\bibliography{pedbib} 

\clearpage

\newpage

\renewcommand{\thetable}{S\arabic{table}}
\renewcommand{\thefigure}{S\arabic{figure}}
\renewcommand{\thesection}{S}
\renewcommand{\thesubsection}{S\arabic{subsection}}
\setcounter{section}{0}

\markboth{{Publication A Forward propagation of a push through a row of people}}{{Supplementary Information}}
\section{Supplementary Information}
\markboth{{Publication A Forward propagation of a push through a row of people}}{{Supplementary Information}}

 \setcounter{subsection}{0}
 \setcounter{table}{0}
  \setcounter{figure}{0}

\subsection{Pressure measurement}
\label{appendix:pressure}

\begin{figure}[htbp]
 \centering
\includegraphics[width=\textwidth]{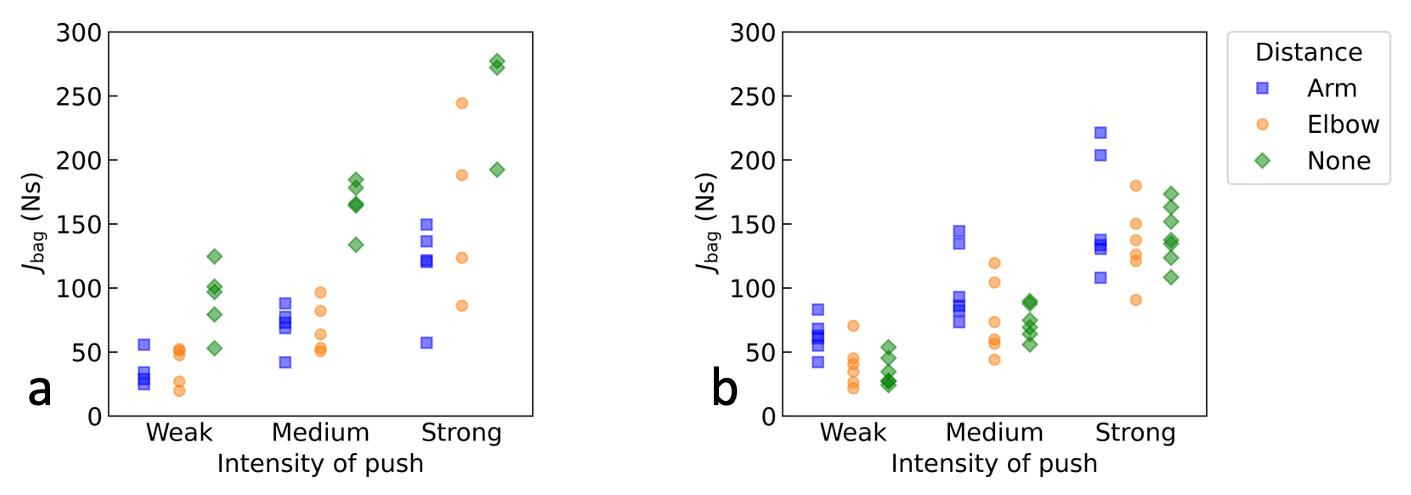}
\caption{Impulse of the push measured at the punching bag for (a) experiments without wall and (b) experiments with wall. There is a general correlation between the perceived strength and the recorded values, but the impulse is affected by the initial inter-person distance as well as the experimental condition (with or without wall).}
\label{fig:pressure_bag}
\end{figure}

\begin{figure}[htbp]
 \centering
\includegraphics[width=0.5\textwidth]{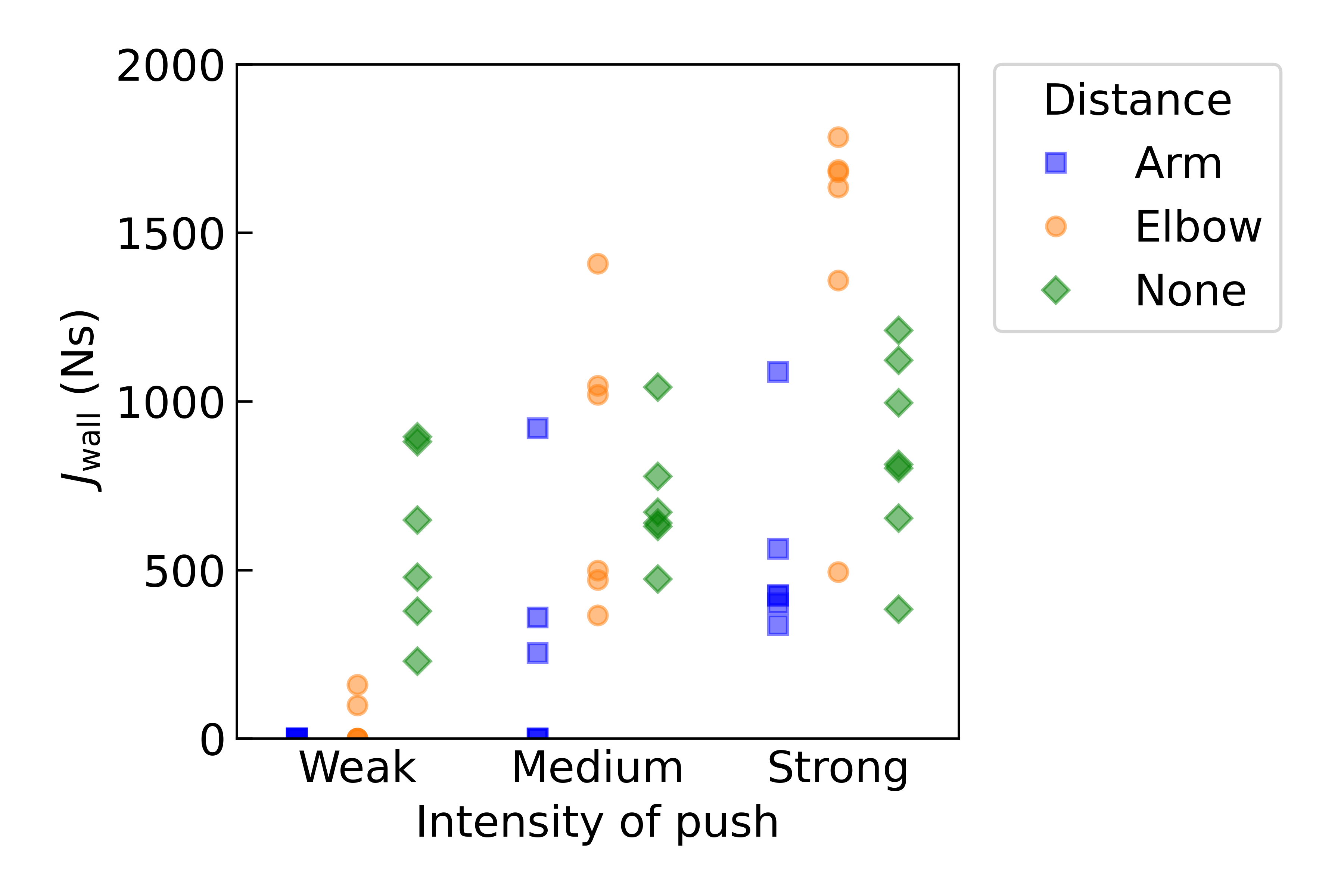}
\caption{Impulse measured at the wall compared to perception of push.Please note: Data recorded at the wall cannot be quantitatively compared to data recorded at the punching bag.}
\label{fig:pressure_wall_b}
\end{figure}

The measured impulse of each push, which is differentiated according to the initial inter-person distances, is compared with the three categories (weak, medium and strong) based on the perception of intensity for the first pushed person in Figure \ref{fig:pressure_bag}.
Thereby, the experimental series with wall (Figure \ref{fig:pressure_bag}\,b) and without (Figure \ref{fig:pressure_bag}\,a) are considered separately.

Within the same perceived strength, the measured impulses show a large variation, comparable to the range between the different categories (weak, medium, strong).
Although there is a general correlation between the perceived strength and the recorded values, it is noticeable that the initial inter-person distance as well as the experimental condition (with or without wall) have an influence.
In the experiments without a wall the values for no distance are often larger than for elbow or arm distance, whereas in the experiments with a wall it is vice versa.
We conclude that the participants' perception is not a good measure for estimating the strength of an impact.
Therefore, the measured impulses are used for further analysis.
\\

The same analysis as well as the comparison of calculated impulse to the three categories is conducted for the pressure values measured at the wall (Figure \ref{fig:pressure_wall_b}).
In case of weak pushes, there is no impact on the wall at arm distance and only a small impact at elbow distance. 
For all other conditions, the impulse data has a high scatter.
When the participants were standing with no distance to each other, there is no significant difference in the calculated impulses among the three perceptions.

\subsection{Moderation analysis}
\label{appendix:statistics1}

A moderation analysis was conducted in R version 4.3.2. The results are shown in Table \ref{tab:statistics} for the comparison of arm postures up, free and down. 

\begin{table}[htbp]
\centering
    \caption{Moderation analysis comparing arms up and arms free to arms down}
    \label{tab:statistics}
    \begin{tabular}{lrrrll} 
    	\toprule
	\addlinespace
	 Residuals: &  \multicolumn{1}{l}{Min} &  \multicolumn{1}{l}{1Q} &  \multicolumn{1}{l}{Median} &  \multicolumn{1}{l}{3Q} &  \multicolumn{1}{l}{Max} \\
	 & \multicolumn{1}{l}{-0.8591} & \multicolumn{1}{l}{-0.2509} & \multicolumn{1}{l}{0.0002} & \multicolumn{1}{l}{0.2561} & \multicolumn{1}{l}{1.1630} \\
	 \addlinespace
	\toprule
	\addlinespace
        Coefficients &  \multicolumn{1}{c}{Estimate} &  \multicolumn{1}{c}{Std. Error} &  \multicolumn{1}{c}{t}  &  \multicolumn{1}{c}{p} &Signif. \\
         \midrule
	intercept             & 1.1303  & 0.1701 &  6.644 & 2.3e-09 & *** \\
	impulse of push                 & 0.0121  & 0.0018 &  6.695 &1.8e-09 & *** \\
	arm posture free &         0.2296 &  0.2319 &  0.990 & 0.32476 &  \\  
	arm posture up    &        0.8025  & 0.2328  & 3.447 & 0.00086 & *** \\
	impulse : arm posture free & -0.0011 &  0.0025 & -0.437 & 0.66305 & \\   
	impulse : arm posture up  & -0.0041  & 0.0021 & -1.944 & 0.05502 & . \\
	\midrule
	\addlinespace
	\multicolumn{5}{l}{Signif. codes: 0 ‘***’ 0.001 ‘**’ 0.01 ‘*’ 0.05 ‘.’ 0.1 ‘ ’ 1} \\
	\addlinespace
	\toprule
	\addlinespace
	\multicolumn{5}{l}{Residual standard error: 0.4288 on 90 degrees of freedom} \\
	\multicolumn{5}{l}{Multiple R-squared:  0.6952,	Adjusted R-squared:  0.6782 } \\
	\multicolumn{5}{l}{F-statistic: 41.05 on 5 and 90 DF,  p-value: $<$ 2.2e-16} \\
	\addlinespace
	\bottomrule
    \end{tabular}
\end{table}

\end{document}